# Development of a Conversation State Prediction System

**Sujay Uday Rittikar**
SUJ00RIT20@GMAIL.COM
*DKTE's Textile and Engineering Institute*
*Maharashtra, IN 416115*

**Abstract**

With the evolution of the concept of Speaker diarization using LSTM, it's relatively easier to understand the speaker identities for specific segments of input audio stream data than manually tagging the data. With such a concept, it's highly desirable to consider the possibility of using the identified speaker identities to aid in predicting the future Speaker States in a conversation. In this study, the Markov Chains are used to identify and update the Speaker States for the next conversations between the same set of speakers, to enable identification of their states in the most natural and long conversations. The model is based on several audio samples from natural conversations of three or greater than three speakers in two datasets, with overall total error percentages for recognized states being lesser than or equal to 12%. The findings imply that the proposed extension to the Speaker diarization is effective to predict the states for a conversation.

**Keywords:** Conversational AI, Sound, Audio Signal Processing, Automatic Speech Recognition, Speaker Recognition, Conversation States, LSTM, Markov Chains

## 1. Introduction

Speaker Recognition (Zhongxin Bai, et al., 2020) is a process to recognize the identity of a speaker as well as verify the identity. Although, the scopes of Speaker identification and Speaker verification are different and may merge based on the applications. Speaker identification has evolved with several text-dependent methods like HMM-Based Methods and text-independent methods such as the VQ-Based approach (Lei Z., et al., 2005) followed by classification using SVM. Speaker diarization(Quan Wang, Carlton Downey, Li Wan, Philip Andrew Mansfield, Ignacio Lopez Moreno, 2017) is a process used in order to perform Speaker identification using the process of partitioning an input audio stream into homogeneous segments according to the speaker identity. According to the findings of Speaker diarization, it's possible to identify the multiple speaker states in an audio stream.

The speaker diarization system is based on the use of Audio embeddings in form of text-independent d-vectors(Jung, J., et al., 2018) to train the LSTM-based (Sepp Hochreiter and Jürgen Schmidhuber, 1997) speaker verification model and furthermore, combine the model with a spectral clustering algorithm to obtain the text-independent states. Thus, the study in the paper is based on the speech audio datasets having firmly speaker voices with minimum noise levels taking into consideration the possible noises to be detected as individual speaker states.

In recent years, Conversational AI (P. Kulkarni, et al., 2019) is used in various applications such as Chatbots and other virtual assistants. It works on the principles of Natural Language Processing (Khurana, et al., 2017) and is dependent on a set of steps essential to identify, retrieve and predict the characteristics of a conversation. The concepts of preprocessing NLP techniques such as tokenization and lemmatization, intent classification, entities extraction, featurizer, and response selector are based on various text-related and vector-based operations and, Machine Learning and Deep Learning-based models such as SVM, Bayesian Networks (Weissenbacher, Davy, 2006), and LSTM.

The Conversational AI richly deals with intent identification and its usage to recognize the flow of conversation and response selection. Although, with advancements in such AI systems

(Sungdong Kim, et al., 2019), the speakers are no more limited to the speaker and the system. The count of speakers in such conversations is higher than two and may cause a problem in selecting the response for a specific user on the basis of intents alone. The values of intents may overlap for multiple speakers at one instance and the response selector may fail to recognize the correct response for multiple speakers due to multiple reasons including the problem of speed to identify the response as well as the correctness of identifying the speaker to send the response to. The problem focuses on creating a Conversational AI system that can converse with multiple users at the same time with no difficulties in identifying the correct responses.

In this system, we initially use the process of Speaker diarization to identify the speaker states in a conversation and use the states to find the probable speaker states for the next conversation holding account of the same speakers. This will improvise the Conversational AI systems to identify the speaker states of a future conversation in advance and thus, would enable it to identify the intents and flow of conversation, thus regulating a faster response rate to each user in the conversation.

## 2. Speaker Diarization with LSTM

### 2.1 Data Collection

The conversations in audio data used in this study are based on natural conversations to ensure that we obtain the practicality of this system to a higher extent. The audio samples for each dataset are collected through the processing of conversations from live videos at the standard sample rate of 44.1 kHz, each of length varying between 5 to 9 minutes. The samples are organized in two datasets representing two different conversations. Each conversation has to be considered using several sample files to form a relevant transition matrix after checking the errors, after generating the transition matrix for the successive sample. The other conversation audio files tested upon are considered as sample phonetic conversations obtained from standard public datasets.

### 2.2 Speech Detection and Segmentation

For inspecting the presence of speech in the Audio Data, Voice Activity Detection (VAD) (Thomas Drugman, et al., 2019) is used. It distinguishes speech segments from background noise in an audio stream. This study segregates the speech segments from non-speech segments in the audio sample files. It is performed to trim out silences and non-speech parts from speech audio data. The Voice Activity Detection system is based on using the filter-based features using the Mel Frequency Cepstral Coefficients(MFCCs) (P. M. Chauhan and N. P. Desai, 2014) representations and source-related features consisting of log-energy and zero-crossing rate(ZCR) features (Goswami, et al., 2013), derived from the power spectral density as an input of a single-layer ANN-based classifier. Furthermore, using a VAD, we obtain the speech segments from the audio which are further divided into smaller non-overlapping segments.

### 2.3 Embedding Extraction

Several speaker verification systems have evolved for different subjects of text-independent and text-dependent systems. Although the embedding is based on a text-dependent speaker verification system, with the advancement of Generalized End-to-End Loss for Speaker Verification, we can obtain the d-vectors (Quan Wang, et al., 2017) for text-independent speaker verification. The embedding extraction is performed using the d-vector feature space over the combination of i-vectors (Ville Vestman, et al., 2020) and PLDA. The d-vector feature extraction is based on supervised DNN architecture as a speaker feature extractor to classify the speakers.

Target labels in this architecture are based on a 1-hot N-dimensional vector where the only non-zero component is the one corresponding to the speaker identity.

## 2.4 Clustering

The clustering algorithm considered in the Speaker diarization system is Spectral offline clustering (Quan Wang, et al., 2017). Offline clustering ensures that the speaker labels are produced after the embeddings of all segments are available thus, the additional contextual information is not ignored. The algorithm consists of the construction of an affinity matrix A and proceeding with a sequence of refinement operations consisting of Gaussian Blur (P. Singhal, et al., 2017), Row-wise Thresholding, Symmetrization, Diffusion, and Row-wise Normalization on the affinity matrix to obtain the eigenvectors. These are further clustered using the K-Means algorithm to produce the speaker labels.

## 3. Markov Chains

After obtaining the speaker states for specific intervals, it's important to note their sequences to understand the further scope of Conversation State Prediction.

The Conversation state prediction refers to the process of identifying the flow of conversation on the basis of speaker state data. It can be achieved using Markov Chains (Chan, et al., 2012) since they are effective in identifying the uncertainties of sequences and provide better recognition using transition probabilities of deterministic states.

The Conversation State Prediction refers to the process of identifying the flow of conversation on the basis of speaker state data. It can be achieved using Markov Chains since they are effective in identifying the uncertainties of sequences and provide better recognition using transition probabilities of deterministic states.

The mechanism is initiated with Speaker states Recognition for some particular number of instances of sequences. By feeding the predicted speaker states in the form of their probability transition matrix to Markov Chains, we provide a pathway to recognize the states in the next instances without any particular need of the process of Speaker states Recognition every time.

Later on, the process of Speaker states Recognition will be performed only when the occasional checks of our system will detect anomalies in recognized states with respect to the actual states. The checker will be run at some specific or random intervals of instances processed by Markov Chains. Repetition of the process will update the Markov Chains with the newly recognized states rather than considering the transition probabilities of previous states. In case there is no anomaly found, the Markov Chains will update their transition matrix at each instance of sequence, making it more reliable.

The Evaluation methods will be performed dynamically using the checker to reciprocate further to Markov Chains whenever the Speaker State Recognition process has to jump in for improving the overall state recognition. This ensures minimal process calls and maximal improvement.

Along with such an advantage in Speaker Recognition, the Markov Chains ensure a predetermined trigger of Conversation states for Conversational AI systems. This can lead to an assumption of possible intents considering a particular speaker state beforehand. The pre-determination of speaker intents in a conversation can regulate a higher response rate, faster API calls or channel connections as well as regulating the pre-setup required to fill specific entities.

**Algorithm to calculate Transition matrices:**

TRANSITION(StateSequence)
INPUT: Sequence string of Speaker States StateSequence
OUTPUT: Multi-dimensional Transition Matrix representing Transition Probabilities of Speaker states
PossibleStates := [0, 1, 2, ... n]; // All possible Speaker states
TransitionMatrix := []; // An Empty List
begin
st := 0;
repeat
    TemporaryList := []; // An Empty List
    TotalCount := StateSequence.count(st); // Count of all instances of state st in S
    st1 := 0;
    repeat
        if st==st1
            otherStates := list(set(PossibleStates) - set(st));
            // Get all states except st
            probs := 0;
            c :=0;
            repeat
                probs += StateSequence.count(st + otherStates[c])/TotalCount;
                // st + otherStates[c] represents the subsequence
                // for ex: "01", "02"
            until c!= length(otherStates);
            TemporaryList.append(1 - c); // The probability except other states
        else
            TemporaryList.append(StateSequence.count(st + st1));
            // st + st1 represent a subsequence where st!=st1
    until st1 != length of PossibleStates;
    TransitionMatrix.append(TemporaryList)
until st!=length of PossibleStates;
return TransitionMatrix;
end

## 4. Experiment and Evaluation

### 4.1 Overview of Experiment

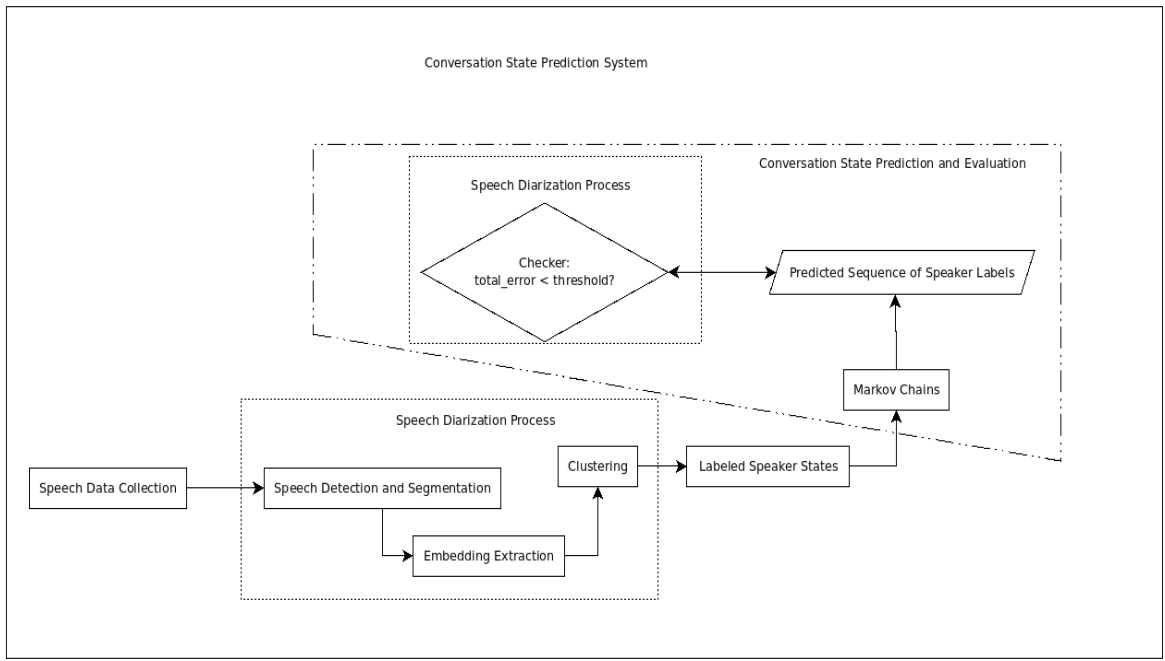

Figure 1: Block Diagram of the Conversation State Prediction system

The Experiment's process starts with Speech Data Collection, passing through the Speech Diarization process as a whole consisting of various sub-processes. Further, the labeled speaker states obtained are used to form a Transition matrix, and thereafter, a Markov chain uses the Transition probabilities to predict a sequence of speaker labels. For the initial iterations, find the best sequence having the least TPE as well as an affordable EPPS. For each iteration, a checker process is initiated. The Checker process uses an instance of the Speech Diarization process and compares the predicted speaker state labels with the actual ones to evaluate total percentage error (TPE) and error percentage per state(EPPS). If TPE is less than 20%(tpe_threshold) and EPPS is less than 30%(epps_threshold), we accept the predicted sequence of states. Otherwise, we will use the labels generated by the Speech Diarization instance used by the Checker process.

### 4.2 Evaluation Metrics

**TPE (Total Percentage Error)**
It is the percentage of incorrect state labels for the entire sequence of predicted state labels.

$$TPE = \frac{100 \times (total\ incorrectly\ predicted\ state\ labels)}{(total\ state\ labels \in sequence)}$$

Figure 2: Equation of Total Percentage Error (TPE)

**EPPS (Error Percentage per State)**
It is the percentage of incorrect state labels for occurrences of each individual state in the sequence of predicted state labels.

$$EPPS = \frac{100 \times (incorrect\ state\ labels\ for\ state\ x)}{(total\ occurrences\ of\ state\ x \in sequence)}$$

Figure 3: Equation of Error Percentage per State (EPPS)

## 5. Results

| Audio File | Speaker State | EPPS (in %) | TPE (in %) |
|:---:|:---:|:---:|:---:|
| 1 | 0 | 8.57 | 9.58 |
|   | 1 | 7.14 |   |
|   | 2 | 20 |   |
| 2 | 0 | 0 | 7.27 |
|   | 1 | 11.54 |   |
|   | 2 | 25 |   |
| 3 | 0 | 0 | 11.11 |
|   | 1 | 25 |   |
|   | 2 | 12.5 |   |
| 4 | 0 | 3.49 | 2.79 |
|   | 1 | 1.09 |   |
|   | 2 | 4.11 |   |
| 5 | 0 | 0 | 3.67 |
|   | 1 | 10 |   |
|   | 2 | 6 |   |
| 6 | 0 | 33.33 | 12 |
|   | 1 | 0 |   |
|   | 2 | 10 |   |
| 7 | 0 | 11.11 | 7.69 |
|   | 1 | 0 |   |
|   | 2 | 13.64 |   |

Table 1: Results on a sample data, in EPPS and TPE

The table illustrates evaluation results of one sample data having 8 audio files with 3 speakers. The first audio file is used as an initial sequence of labels to obtain a transition matrix. The results suggest low TPEs, not greater than 12%. Although, there is an instance where the EPPS is as high as 33.33% and thus, at such instances, the Speech Diarization process should be used to revise those sub-sequences.

## 6. Evaluation and improvement of efficiency

The system maintains a low error rate with most of the data, but we can't ignore the possibility of the predictions being wrong on the basis of how static the transition matrix remains. The transition matrix doesn't change much in terms of its probability values for a single sequence of speaker states. In addition, there is a possibility that it becomes difficult to decide the next state because there is very little difference between the probability values of possible states from a certain state. Therefore, it is tough to make a decision on which state will be preferred the next on the basis of the transition matrix on the entire sequence of data.

In order to solve this problem, we need to:

1. Consider the Transition matrix of a certain number of states or a subsequence to form a basis to predict the next consecutive states.
2. Ease the decision-making process by considering the correct probabilities where, for each state, there is not much confusion to choose from.

Previous Assumption before performing these operations:

We use a window of k speaker states as a single sequence each time, provided k<n where n is the number of total speaker states in a single sequence.

### 6.1 Difference Thresholds

There are 2 major types of differences to be considered before we proceed with the next set of classifications:

a. Difference between the current Transition matrix based on a sequence of n speaker states and Transition matrix based on a windowed sequence/subsequence of k speaker states. We'll refer to this as matrix_diff.
b. Difference between the largest probability in each row i of the Transition matrix and each element of every row j except the element that represents the largest probability. We'll refer to this as row_diff.

These differences have 2 threshold assumptions to be considered respectively as conditions:

a. For the difference between curr. Transition matrix and wind. Transition matrix, there should be no more than 10-15% of error between each element of the Transition matrix. Thus, each element of matrix_diff <= 0.15. The reason behind this assumption is to avoid any major differences between the speaker state sequences to be predicted and those considered to form the predictions.
b. For the difference between the maximum element of a row(row_diff) and each other element of the row, there has to be more than 1/s or 100/s % difference where s refers to the total number of speaker states(identities) in the transition matrix. This assumption simply relies on avoiding similarities between the probabilities in a single row, to avoid a false positive while making the decision of the next speaker state.

### 6.2 The Evaluator

After finding the thresholds, if the conditions for both difference thresholds are satisfied, we have no anomaly to proceed with the next subsequence or a window of k states to check upon and form a transition matrix.

Although, if any of the conditions from 6.1 are not satisfied, it is necessary to form our basis of predicting the next speaker state/s using a certain pattern of speaker states.

To form such a pattern, there are 2 methodologies we can work on:

1. The previous window of sequences based approach
2. Forecast using a pattern of sequences with no uniform time intervals

## 7. Discussion

The scope of multi-user conversational AI will be adapted soon as the Chatbots have started evolving to an extent where they will be reaching Level 4 and Level 5, making them automated as well as adaptive assistants. While the Conversational AI still focuses on the development of a single user to single system approach, we could use multiple subsystems working in parallel on a single system for a single user, making it a centralized conversation system with multiple speaker states. The reverse scenario is also a possibility, considering multiple users interacting with an AI assistant, the proposed system will ensure proper management of such a complex conversation.

The further challenges for this model of the system stand as regulating the state labels at the instances when the error percentages are high. The current methodology of changing the states stands dispensable as it'll replace the whole sequence. Thus, a method to change the particular states without over-optimizing the errors will be convenient to get better results.

While the applications with the current proposition of this model don't range vastly, the concept of developing a separate Conversation State Prediction system is a highly valid concept in the current evolving Conversational AI. While the LSTM alone are amazing to use, the requirement of such a system individually can not be avoided since the current AI systems are going to advance to a position where there will be several agents communicating but no specific system to handle the internal traffic of Conversation. Recognizing the sequence of conversation states can enable such systems to maintain the flow without any interruptions. In addition, the concepts of intents and entities can also be improved as the system will maintain its usage of architectures such as DIET(Bunk, Tanja & Varshneya, Daksh & Vlasov, Vladimir & Nichol, Alan, 2020) on the basis of the future intents known in advance.

The other applications also include building predictive movies and entertainment systems. While there are multiple possibilities, it is important to derive such system architectures using more complex models than the proposition in this paper of using Markov Chains. The simplistic applications can, though, use the Markov Chains or even HMM to recognize the speaker states as discussed in this paper with extensions of Sentiment Analysis in order to .

**Acknowledgments**

Many thanks to Prajwal Hukerikar for helpful discussions and Apoorv Terwadkar for providing the Data in the form of recordings for the study.